\documentclass[aps,prl,twocolumn,groupedaddress]{revtex4}
\usepackage{amsfonts}
\usepackage{mathrsfs}
\usepackage{graphicx}
\usepackage{subfigure}
\usepackage{amsmath}
\usepackage{amssymb}
\usepackage{amsbsy}
\usepackage{bm}

\def\bs{\bm{\sigma}}
\def\bx{\bm{\xi}}
\def\bxtF{\bm{\xi}^{1,{\rm true}}}
\def\bxtS{\bm{\xi}^{2,{\rm true}}}
\def\bxF{\xi^{1,{\rm true}}}
\def\bxS{\xi^{2,{\rm true}}}
\def\gmba{\Gamma_{b\rightarrow i}^1}
\def\gmbb{\Gamma_{b\rightarrow i}^2}
\def\gba{G_{b\rightarrow i}^1}
\def\gbb{G_{b\rightarrow i}^2}
\def\Xb{\Xi_{b\rightarrow i}}
\def\xab{\xi_i^1\xi_i^2}

\begin{document}

\title{Statistical physics of unsupervised learning with prior knowledge in neural networks}

\author{Tianqi Hou}
\affiliation{Department of Physics, the Hong Kong University of Science and Technology, Clear Water Bay, Hong Kong, People's Republic of China}
\author{Haiping Huang}
\email{huanghp7@mail.sysu.edu.cn}
\affiliation{PMI Lab, School of Physics,
Sun Yat-sen University, Guangzhou 510275, People's Republic of China}

\date{\today}

\begin{abstract}
Integrating sensory inputs with prior beliefs from past experiences in unsupervised learning is 
a common and fundamental characteristic of brain or artificial neural computation.
However, a quantitative role of prior knowledge in unsupervised learning remains unclear, prohibiting a
scientific understanding of unsupervised learning. Here, we propose a statistical physics model of unsupervised learning with
prior knowledge, revealing that the sensory inputs drive a series of continuous phase transitions related to
spontaneous intrinsic-symmetry breaking. The intrinsic symmetry includes both reverse symmetry and permutation symmetry, commonly observed in most artificial neural networks.
Compared to the prior-free scenario, the prior reduces more strongly
the minimal data size triggering the reverse symmetry breaking transition, and moreover, the prior merges, rather than separates, permutation symmetry breaking phases. 
We claim that the prior can be learned from data samples, which in physics corresponds to a two-parameter Nishimori constraint.
This work thus reveals mechanisms about the influence of the prior on unsupervised learning.
\end{abstract}

 \maketitle

The sensory cortex in the brain extracts statistical regularities in the environment in an unsupervised way.
This kind of learning is called unsupervised learning~\cite{Barlow-1989}, 
relying only on raw sensory inputs, thereby thought of as a fundamental function of the sensory cortex~\cite{Marr-1970}.
When sensory information is uncertain, a natural way to model the outside world is integrating sensory inputs with internal
prior beliefs, in accordance with Bayesian inference. Bayesian theory formalizes how the likelihood function of sensory inputs and
the prior statistics can be coherently combined, taking a trade-off between input feature reliability, as encoded in the likelihood function, and 
the prior distribution~\cite{Kersten-2004}. The Bayesian brain hypothesis~\cite{Bayes-2016}
was widely used to model sensorimotor behavior~\cite{Kord-2004,motor-2018}, perceptual decision making~\cite{Beck-2008},
object perception in visual cortex~\cite{Alan-2006,Lee-2003}, and even cognition~\cite{Human-2011}. Bayesian inference is also a principled 
computational framework in deep-learning-based machine learning~\cite{Blun-2015,Pbp-2015,Welling-2018}. When the prior beliefs are 
taken into account, the unsupervised learning meets Bayesian inference. Thus incorporating prior beliefs from past experiences into
unsupervised learning is a common and fundamental characteristic of the computation in the brain or artificial neural networks.
However, current studies mostly focused on
neural implementations of the Bayesian inference~\cite{Bayes-2016,Bayes-2019}, or focused on designing scalable Bayesian learning algorithms for deep networks~\cite{Pbp-2015,Welling-2018},
making a scientific understanding of unsupervised learning with prior knowledge lag far behind its neural implementations or engineering applications.

By asking a minimal data size to trigger learning in a two-layer neural network, namely a restricted Boltzmann machine (RBM)~\cite{Hinton-2006a},
a recent study claimed that sensory inputs (or data streams) are able to drive a series of phase transitions related to broken inherent-symmetries of the model~\cite{Huang-2019}.
However, this model does not assume any prior knowledge during learning, therefore the impact of priors on the learning remains unexplained.
Whether a learning is data-driven or prior-driven depends highly on feature reliability of sensory inputs. When sensory inputs become
highly unreliable, prior knowledge dominates the learning. Otherwise, the data likelihood takes over. A quantitative role of prior knowledge is thus
a key to unlock the underpinning of unsupervised learning with priors.

Here, we propose a mean-field model of unsupervised learning with prior knowledge, to provide an analytical argument supporting 
surprising computational roles of priors. First, the prior knowledge reduces strongly the minimal data (observations) size at which
the concept-formation starts, as \textit{quantitatively} predicted by our model. Second, phase transitions observed in the RBM model of unsupervised 
learning without priors are significantly reshaped, showing that the prior shifts an intermediate phase observed in the model
without priors. Lastly, our theory reveals that the variability in data samples, as encoded by a temperature-like hyper-parameter,
as well as the intrinsic correlation between synapses connecting layers of the RBM, can be learned directly from the data.
In physics, this corresponds to a two-parameter Nishimori constraint, generalizing the original concept of a single-parameter
Nishimori line~\cite{Nishimori-2001}. Therefore, our model provides deep insights about roles of prior knowledge in unsupervised learning.

From a neural network perspective, 
the interplay between the prior and the likelihood function of 
data can be captured by synaptic weights. These synaptic weights are modeled by feedforward connections
in a RBM~\cite{Hinton-2006a,Huang-2017,Huang-2019}. More precisely, the RBM is a two-layer neural network where
there do not exist intra-layer connections. The first layer is called the visible layer, receiving sensory inputs (e.g., images),
while the second layer is called the hidden layer, where each neuron's input weights are called the receptive field (RF) of that hidden neuron.
In an unsupervised learning task, these synaptic weights are adjusted to
encode latent features in the data.  
We assume that
both the neural and synaptic states take a binary value ($\pm1$). The RBM is a universal approximator of any discrete distributions~\cite{Bengio-2008}. For simplicity, 
we consider only two hidden neurons. Therefore, the joint activity distribution reads as follows~\cite{Hinton-2002,Huang-2015b},
\begin{equation}\label{Pvh}
 P(\bs,h_{1},h_{2}|\bx^{1},\bx^{2})=\frac{1}{Z(\bx^{1},\bx^{2})}e^{\frac{\beta}{\sqrt{N}}\sum_{j}h_j\sum_{i\in\partial j}\xi^{j}_{i}\sigma_i},  
\end{equation}
where $\partial j$ denotes neighbors of the node $j$, $\bs$ indicates an $N$-dimensional sensory input, $h_1$ and $h_2$ are the hidden neural activity,
an inverse-temperature $\beta$ characterizes the noise level of the input, the network-size
scaling factor ensures that the free energy of the model is an
extensive quantity, finally $\bx^1$ and $\bx^2$ are the receptive fields of the two hidden neurons, respectively.
$Z(\bx^1,\bx^2)$ is the partition function in physics. Another salient feature in this model is that
the hidden activity can be marginalized out, which helps the following analysis.

For an unsupervised learning task, one can only have access to raw data, defined by $\mathcal{D}=\{\bs^a\}_{a=1}^M$. We assume a
weak dependence among data samples~\cite{Hinton-2002}. 
The task is to infer the synaptic weights encoding latent features in the data.
This is naturally expressed as computing the posterior probability, and thus the Bayes' rule applies as follows,
\begin{equation}\label{Ppost}
    \begin{split}
      & P(\bx^1,\bx^2 |\mathcal{D})=\frac{\prod _{a} P(\bs^{a} |\bx^1,\bx^2 ) \prod_{i=1}^{N} P_{0}({\xi_{i}^{1},\xi_{i}^{2}})}
       { \sum_{\bx^1,\bx^2    }  \prod _{a} P(\bs^{a} |\bx^1,\bx^2 ) \prod_{i=1}^{N} P_{0}({\xi_{i}^{1},\xi_{i}^{2}})} \\
      & = \frac{1}{\Omega} \prod_{a} \frac{1}{\cosh{(\beta^{2}Q)}}\cosh{\bigg  (\frac{\beta}{\sqrt{N}}\bx^{1}\cdot\bs^{a}  \bigg )}\\
      &\times\cosh{ \bigg  (\frac{\beta}{\sqrt{N}}\bx^{2}\cdot\bs^{a}\bigg )}\prod_{i=1}^{N}P_{0}(\xi_{i}^{1},\xi_{i}^{2}   ),
    \end{split}
\end{equation}
where $Q=\frac{1}{N}\sum_i\xab$ stemming from $Z(\bx^1,\bx^2)\simeq2^Ne^{\beta^2}\cosh(\beta^2Q)$~\cite{Huang-2019}, $\Omega$ is the partition function for the 
learning process, and $P_0(\xi_i^1,\xi_i^2)=\frac{1+q}{4}\delta(\xi_i^1-\xi_i^2)+\frac{1-q}{4}\delta(\xi_i^1+\xi_i^2)$ specifying
the prior knowledge we have about the distribution of synaptic weights. Note that $q$ determines the correlation level of
the two RFs. The prior $P_0(\xi_i^1,\xi_i^2)$ can be recast into another form of $P_0(\xi_i^1,\xi_i^2)=\frac{e^{J_0\xi^1_i\xi^2_i}}{4\cosh(J_0)}$ where
$J_0=\tanh^{-1}q$. The unsupervised learning can thus be investigated within a teacher-student scenario. First, we prepare a teacher-type RBM whose synaptic weights
are generated from the prior distribution with a prescribed $q$. Then the data $\mathcal{D}$ is collected from the equilibrium state 
of the model (Eq.~(\ref{Pvh})) through Gibbs sampling~\cite{Hinton-2002}. Finally, a student-type RBM tries to infer the teacher's RFs only based on
the noisy data the teacher generates with another prescribed $\beta$. Therefore, the unsupervised learning with prior knowledge (including both $q$ and $\beta$)
can be studied within the Bayes-optimal framework (Eq.~(\ref{Ppost}))~\cite{Iba-1999}. Note that there exist two types of inherent symmetries in the model, i.e.,
the model (Eq.~(\ref{Ppost})) is invariant under the operation of $\bx\rightarrow-\bx$ for two hidden nodes (reverse symmetry)
or the permutation of $\bx^1$ and $\bx^2$ (permutation symmetry (PS)).

The teacher-student scenario is amenable for a theoretical analysis, since the data distribution can be analytically calculated.
This is different from numerical experiments on a real dataset whose exact distribution is not accessible. Moreover, we are interested in
the limits---$M\rightarrow\infty$ and $N\rightarrow\infty$ but the ratio or data density is kept constant as
$\alpha=\frac{M}{N}$.

The emergent behavior of the model is captured by the free energy function defined by $-\beta Nf=\left<\ln\Omega\right>$,
where $\left<\bullet\right>$ denotes the disorder average over both the distribution of
planted true RFs and the corresponding data distribution $P(\mathcal{D}|\bx^1,\bx^2)$.
However, a direct computation of the disorder average is impossible. Fortunately, by introducing $n$ replicas of the original system,
we can estimate the free energy density by applying a mathematical identity $-\beta f=\lim_{n\rightarrow 0,N\rightarrow\infty}\frac{\ln\left<\Omega^n\right>}
{nN}$ where we only need to evaluate an integer power of $\Omega$. Technical calculations are deferred to Supplemental Material~\cite{SM}. Here we quote only the final result, and 
give an intuitive explanation. For simplicity, we assume that order parameters of the model (defined below) are invariant under permutation of replica indexes.
This is called the replica symmetric (RS) Ansatz in spin glass theory~\cite{Mezard-1987}. 

The order parameters $(T_1,T_2,q_1,q_2,\tau_1,\tau_2,R,r)$ and their associated conjugated counterparts are 
stationary points of the free energy function, obtained through a saddle-point analysis in the thermodynamic limit~\cite{SM}.
The order parameters are calculated as follows,
\begin{subequations}\label{OP}
  \begin{align}
      T_{1}&=[\bxF\langle \xi^{1} \rangle ] , 
      T_{2}=[\bxS\langle  \xi^{2} \rangle ],\\
      q_{1}&=[  \langle  \xi^{1}  \rangle^{2}  ], 
      q_{2}=[  \langle  \xi^{2}  \rangle^{2}  ],  \\
      \tau_{1}&=[ \bxF \langle  \xi^{2} \rangle ] , 
      \tau_{2}=[ \bxS \langle  \xi^{1} \rangle ],  \\
      R&=[\langle \xi^{1}\xi^{2} \rangle], 
      r=[\langle \xi^{1}\rangle \langle \xi^{2} \rangle ],
  \end{align}
\end{subequations}
where $\left<\bullet\right>$ denotes an average under the Boltzmann measure of an effective
two-spin interaction Hamiltonian, arising from the entropic computation of the replica method~\cite{SM}, i.e., $P_{{\rm eff}}(\xi^1,\xi^2)\propto
\frac{1}{4\cosh J_0}e^{b_1\xi^1+b_2\xi^2+(b_3+J_0)\xi^1\xi^2}$ where $b_1$ (or $b_2$) and $b_3$ are  random effective fields and couplings, respectively,
related to conjugated order parameters $(\hat{T}_1,\hat{T}_2,\hat{q}_1,\hat{q}_2,\hat{\tau}_1,\hat{\tau}_2,
\hat{R},\hat{r})$, and
$[\bullet]$ indicates an average over the standard Gaussian random variables and
the true prior $P_0(\bxF,\bxS)$. These order parameters capture the emergent behavior of our model.
$T_1$ and $T_2$ characterize the overlap between prediction and ground truth. $q_1$ and $q_2$ characterize the self-overlap (Edwards-Anderson order parameter in physics~\cite{Mezard-1987}).
$\tau_1$ and $\tau_2$ characterize the permutation-type overlap. $R$ and $r$ characterize the student's guess on the correlation level of the planted RFs.

Interestingly, when $\alpha$ is small, trivial (null values) order parameters except $R$ are a stable 
solution of Eq.~(\ref{OP}), thereby specifying a random guess (RG) phase. As expected, $R$ reflects the prior information, thus being equal to $q$ irrespective of $\alpha$.
In this phase, $\left<\xi^1\right>=\left<\xi^2\right>=0$, the weight thus takes $\pm1$ with equal probabilities,
implying that the data does not provide any useful information to bias the weight's direction. The underlying
rationale is that the posterior (Eq.~(\ref{Ppost})) is invariant under the reverse operation $\bx\rightarrow-\bx$, and this symmetry is unbroken in the RG phase.

Surprisingly, as more data is supplied, the RG phase would become unstable at a critical data density.
By a linear stability analysis~\cite{SM}, this threshold can be analytically derived as
\begin{equation}\label{thr}
 \alpha_c=\frac{\Lambda(\beta,q)}{(1+|q|)(1+|\tanh(\beta^2q)|)},
\end{equation}
where $\Lambda(\beta,q)=\frac{\beta^{-4}}{1+q\tanh(\beta^2q)+|q+\tanh(\beta^2q)|}$ denoting the learning threshold for the prior-free scenario~\cite{Huang-2019}.
In the correlation-free case ($q=0$), the known threshold $\alpha_c=\beta^{-4}$ is recovered~\cite{Huang-2016b,Huang-2017}.
Compared to the prior-free scenario, the prior knowledge contributes an additional factor leading to a further reduction
of the threshold ($\sim60\%$ of the prior-free one for $q=0.3$ and $\beta=1$). Most interestingly,
in the weak-correlation limit, where $q\sim\beta^{-2}$ with a proportional constant $q_0$ in the presence of less noisy data
(large $\beta$), $\alpha_c\beta^4=\frac{1}{(1+|\tanh q_0|)^2}$, which implies that the learning threshold can be lowered down to only $32\%$ of
the correlation-free case for $q_0=1$. This demonstrates that the weak correlation between synapses plays
a key role of reducing the necessary data size to trigger concept-formation, a surprising prediction of the model.

\begin{figure}
\centering
\subfigure[]{
     \includegraphics[bb=5 7 432 278,scale=0.5]{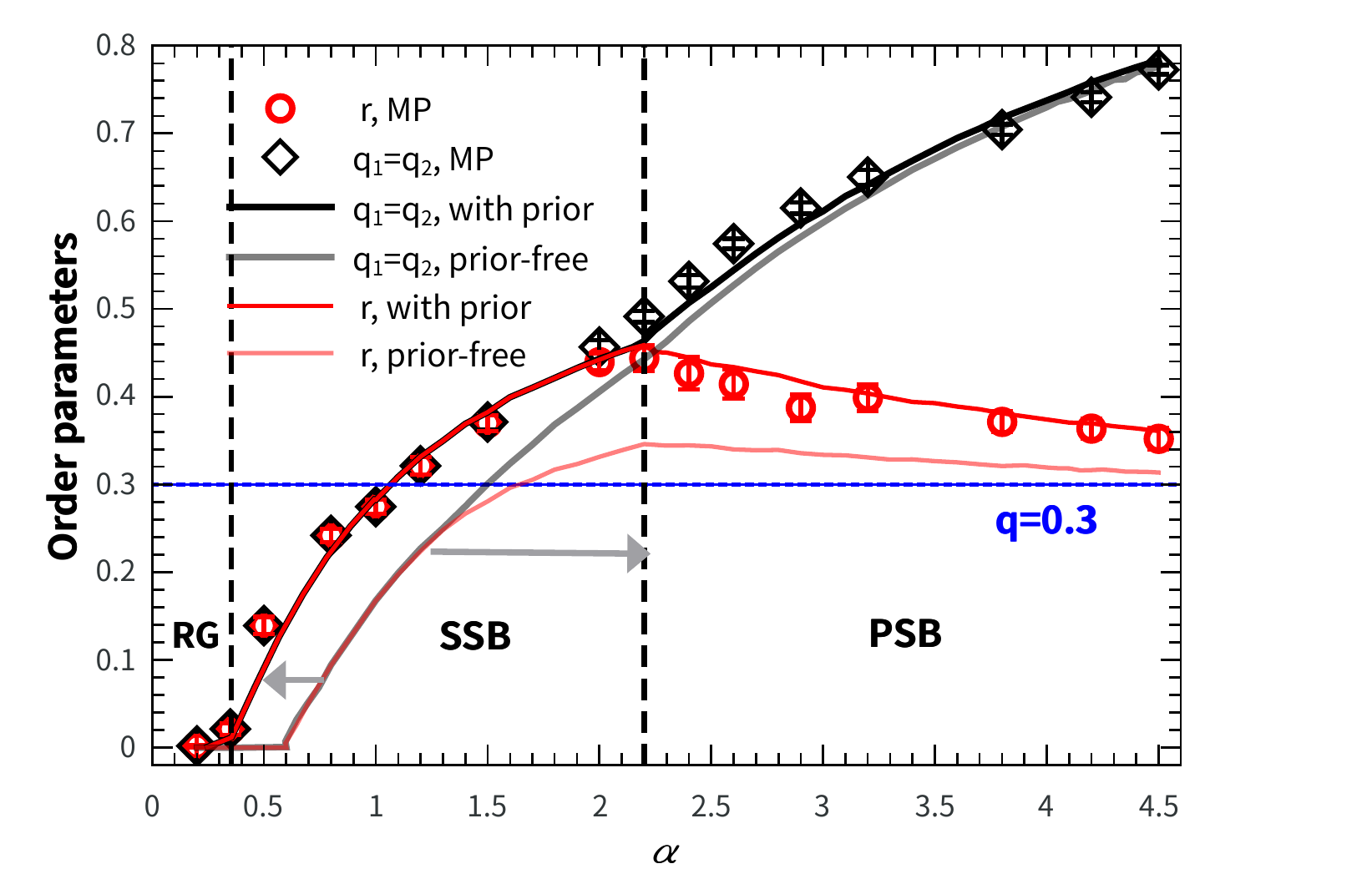}}
     \subfigure[]{\includegraphics[bb=2 9 372 310,scale=0.5]{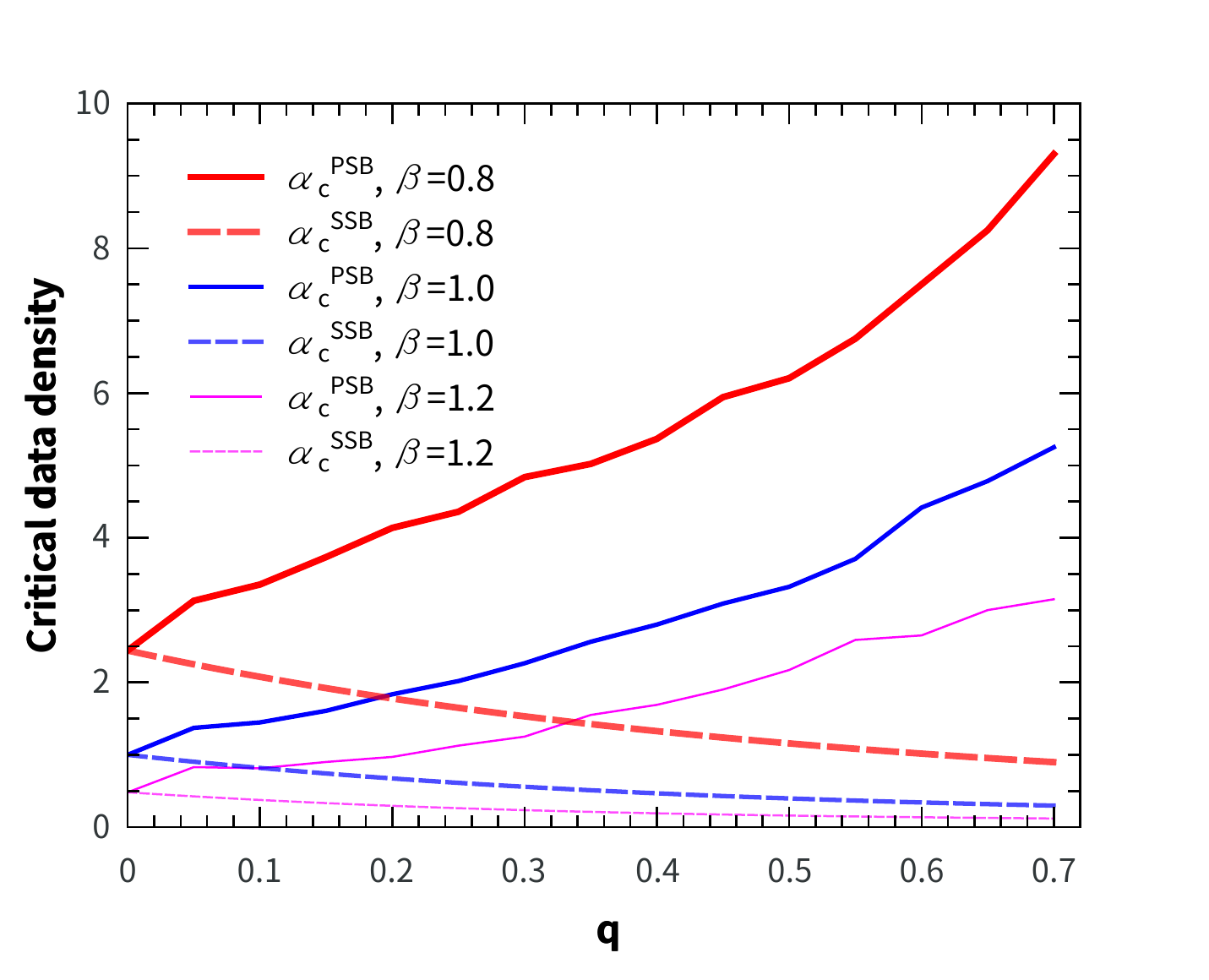}}
  \caption{
  (Color online) Phase diagram of unsupervised learning with priors.
  (a) Order parameters versus data densities with ($\beta,q$)=($1.0,0.3$). Lines are replica results compared with symbols obtained from the message passing (MP) procedure (instances of $N=200$).
  Previous results of the prior-free unsupervised learning~\cite{Huang-2019} are also plotted for comparison. The arrows indicate the role of priors in shifting the phase transition points. 
  (b) Critical data densities for SSB and PSB are obtained from replica analysis and plotted for increasing values of $\beta$.
  }\label{phase}
\end{figure}


When $\alpha>\alpha_c$, the RG phase is replaced by the symmetry-broken phase, where $\left<\xi^1\right>=\left<\xi^2\right>\neq0$ and 
the intrinsic reverse symmetry is spontaneously broken. We thus call the second phase a spontaneous symmetry breaking (SSB) phase.
The SSB leads to a non-zero solution of $q_1=q_2=T_1=T_2=\tau_1=\tau_2=r$. A reasonable interpretation is that,
the student infers only the common part of the two planted RFs. Thus the PS still holds for the student's hidden neurons.
Moreover, $\bxtF$ and $\bxtS$ have the PS property as well, explaining the solution we obtained.
The SSB phase is thus permutation symmetric, which is stable until a turnover of the order parameter $r$ is reached (Fig.~\ref{phase} (a)).

At the turnover, the PS is spontaneously broken, thereby leading to a permutation symmetry breaking (PSB) phase. 
The third phase is characterized by two fixed points: (1) $q_1=q_2=T_1=T_2$, and $\tau_1=\tau_2=r$; (2)
$q_1=q_2=\tau_1=\tau_2$, and $T_1=T_2=r$. These two fixed points share the same free energy, representing two possible choices of ground truth---
$(\bxtF,\bxtS)$ or $(\bxtS,\bxtF)$. In fact, the PSB phase has two subtypes---a ${\rm PSB_s}$ phase where the permutation symmetry between $\bx^1$ and $\bx^2$  
is broken on the student's side, i.e., $\left<\xi^1\right>$ can point conversely to $\left<\xi^2\right>$ yet with the same magnitude, thereby $q_1=q_2\neq r$, and a ${\rm PSB_t}$ phase where the PSB occurs on the teacher's side,
i.e., $\bxtF$ and $\bxtS$ can not be freely permuted, thereby $T_{1,2}\neq\tau_{2,1}$~\cite{Huang-2019}. Interestingly,
the self-overlap deviates from $r$ at the turnover, thereby merging ${\rm PSB_s}$ phase 
and ${\rm PSB_t}$ phase into a single PSB phase, rather than separating these two subtypes as in the prior-free scenario (Fig.~\ref{phase} (a)). 
With the help of prior knowledge, the student is able to distinguish two planted RFs (${\rm PSB_t}$) \textit{at the same time} when starting 
to infer different components of the true RFs (${\rm PSB_s}$). Furthermore, the prior does not change the ${\rm PSB_t}$ transition point of the prior-free case, as knowing $q$ does not help to
accelerate the recognition of two choices of ground truth. However, the knowledge of $q$ does elevate the overlap values before the turnover, leading to a
larger value of $r$ in the post-turnover regime compared to the prior-free case (see a proof in~\cite{SM}).
After the turnover, the overlap equal to $\min(T_1,\tau_1)$ or $\min(T_2,\tau_2)$
has the same value with $r$, since $(\bxtF,\bxtS)$ follows the same posterior as $(\bx^1,\bx^2)$. As expected, $r$ finally
tends to $q$ at a finite but large value of $\alpha$ (Fig.~\ref{phase} (a)).


We conclude that with the prior knowledge, the data stream drives the SSB and PSB phase transitions of continuous type.
Thresholds of the transitions are summarized in Fig.~\ref{phase} (b). This conclusion is verified by
numerical simulations on single instances of the model by applying a message-passing-based learning algorithm (Fig.~\ref{phase} (a)). Briefly,
a cavity probability of $(\xi_i^1,\xi^2_i)$ without considering the $a$-th data sample $P_{i\rightarrow a}(\xi_i^1,\xi_i^2)$ is defined.
Then the cavity magnetization $ m_{i \to a }^{1,2}=\sum_{\xi_{i}^{1},\xi_{i}^{2}}\xi_{i}^{1,2} P_{i \to a}(\xi^1_i,\xi^2_i)   $ and the 
cavity correlation $q_{i \to a}=\sum_{\xi_{i}^{1},\xi_{i}^{2}}  \xi_{i}^{1} \xi_{i}^{2} P_{i \to a}(\xi^{1}_{i},\xi^{2}_{i})$ can be written into
a closed-form iterative equation, using the approximation that the cavity probabilities surrounding a data sample are factorized~\cite{SM}. For evaluating the order parameters, 
the full (not cavity) magnetization $m_{i}^{1,2}$ and correlation $q_i$
can be computed by considering contributions from all data samples. 
\begin{figure}
     \includegraphics[bb=7 12 697 528,width=0.5\textwidth]{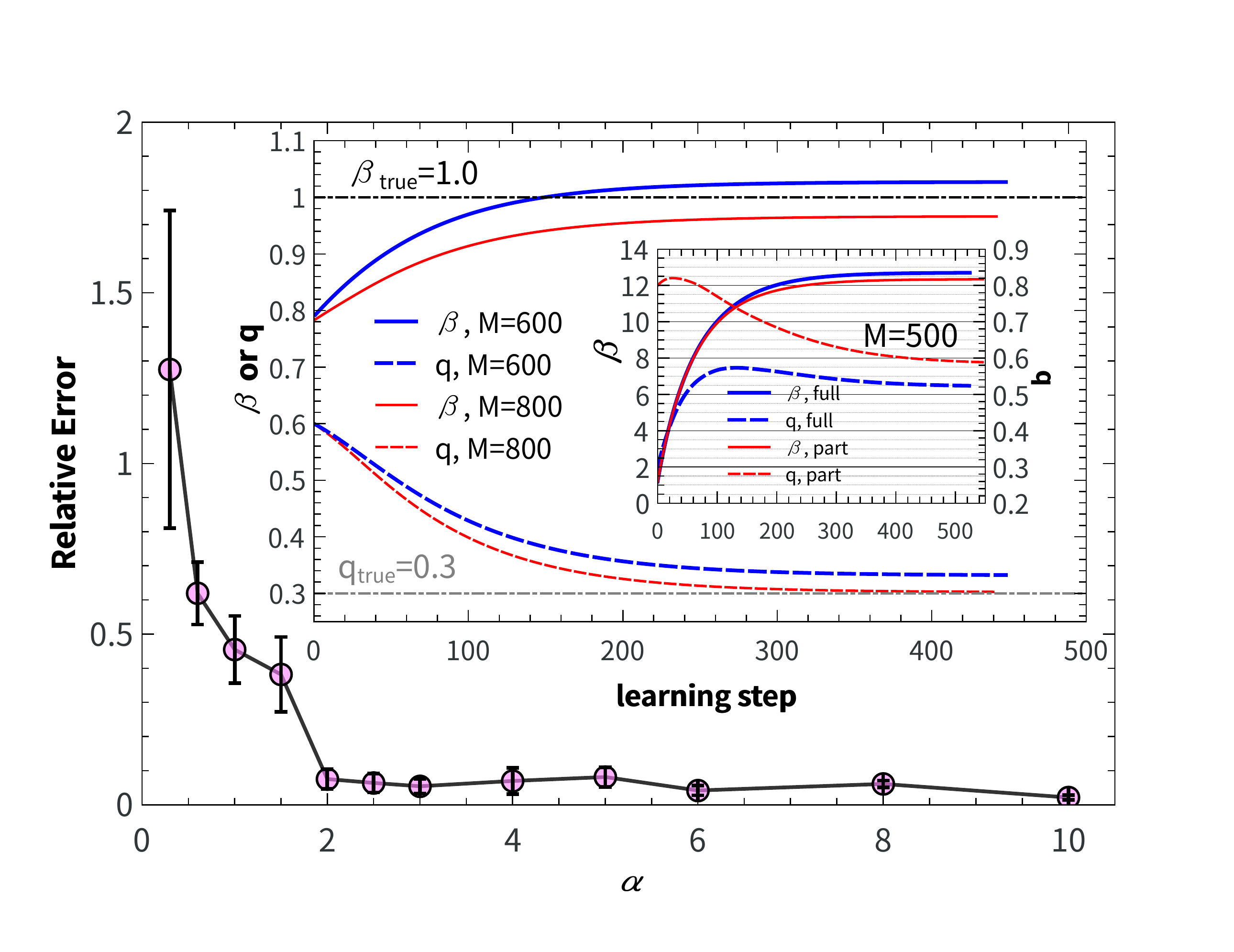}
  \caption{
  (Color online) Inference of the hyper-parameters ($\beta,q$) from raw data. The performance is measured by the relative error defined as
  $|\mathcal{L}_{{\rm est}}/\mathcal{L}_{{\rm true}}-1|$, where $\mathcal{L}$ denotes the negative log-likelihood (Eq.~(\ref{em}))~\cite{SM}. The error bar characterizes the fluctuation among twenty independent runs.
  The inset shows learning trajectories on random instances of the model ($N=100$).
  The sub-inset shows results on the MNIST dataset including full dataset (containing all ten types of digits) or part of the dataset (containing only four types of digits).
  }\label{pred}
\end{figure}

In our model, the learning thresholds related to the SSB and PSB phase transitions are only determined by two 
parameters $\beta$ and $q$. We then ask whether these two parameters can be learned from the raw data (not necessary to explore all 
data samples). In principle, this can be achieved by maximizing the following posterior of the hyper-parameters,
\begin{equation}\label{em}
    P(\beta,q|\mathcal{D})=\sum_{\bx^1,\bx^2}P(\beta,q,\bx^1,\bx^2|\mathcal{D})\propto e^{-\alpha N\beta^2}\Omega,
\end{equation}
where the Bayes' rule is applied~\cite{SM}. The maximum of this posterior has the following property,
\begin{subequations}\label{emeq}
   \begin{align}
       & \beta=-\frac{ \epsilon(\beta ) }{2\alpha},   \\
       &  q=\frac{1}{N}\sum_{i=1}^N q_{i}(\beta,q),
   \end{align}
\end{subequations}
where the energy density $\epsilon(\beta)$ independent of $q$ and the full correlation $\{q_i\}$ can be estimated from the message passing algorithm~\cite{SM}.
Eq.~(\ref{emeq}) constructs a two-parameter Nishimori constraint~\cite{SM}, implying that the energy density of the model with prior knowledge is analytic ($\epsilon=-2\alpha\beta$).
Given only the data samples, although we do not know the hyper-parameters ($\beta$,$q$) underlying the data,
iteratively imposing the Nishimori constraint helps to reach a consistent value of $(\beta,q)$ to explain the data.
In statistics, this iterative scheme is called the expectation-maximization algorithm~\cite{EM-1977}, where the message update to
compute $(\epsilon,\{q_i\})$ is called an expectation-step, while the hyper-parameter update is called a
maximization-step. The hyper-parameter space, especially when the amount of data samples
is not sufficient, is not guaranteed to be convex, instead being highly non-convex in general, as verified by a high relative inference error
with a large fluctuation in a data-deficient regime (Fig.~\ref{pred}).
We first apply this framework to synthetic datasets,
where the hyper-parameters can be accurately predicted provided that the supplied dataset is large enough (Fig.~\ref{pred}).
Using the current neural network model with $N$ input neurons and two hidden neurons, we then apply this algorithm to a handwritten digit dataset~\cite{mnist}, with $N=784$. We obtain a value of $(\beta,q)=(12.697,0.524)$ (Fig.~\ref{pred}).
This predicted $\beta$ value is much smaller than that reported in a one-bit RBM~\cite{Huang-2017}.

\textit{Summary.}---Learning statistical regularities in sensory inputs is naturally characterized by integrating sensory information with
prior beliefs on the latent features. Our theory clarifies the role of prior knowledge, when unsupervised learning meets Bayesian inference.
Incorporating the prior merges ${\rm PSB_{s}}$ and ${\rm PSB_t}$ phases, while the
${\rm PSB_t}$ phase lags behind the ${\rm PSB_s}$ phase in the prior-free model. Thus the prior is able to reshape the concept-formation process in unsupervised learning.
This is one key prediction of our theory. As expected, the prior knowledge reduces much more
significantly the minimal data size triggering SSB than the prior-free case. Moreover, a weak correlation between synapses reduces strongly
the minimal data size as well, in consistence with the well-known non-redundant weight hypothesis~\cite{Barlow-1961}.
This contributes to the second prediction of the model.
The PSB phase was also recently observed in a RBM with more than two hidden neurons~\cite{Huang-2019data}.
The predictions may be generalized to a hierarchical complex system, and even testable in a visual hierarchy where top-down contextual priors 
are combined with bottom-up observations when implementing probabilistic inference~\cite{Lee-2003,Kersten-2004}.

Therefore, using the concept of SSB and PSB in physics, 
our theoretical study provides deep insights about roles of prior knowledge 
in unsupervised Bayesian learning, which is linked to the Bayesian brain hypothesis (e.g., when sensory inputs become highly unreliable, 
prior knowledge dominates the learning) and the Bayesian inference in neural networks 
(both prior-induced SSB and PSB are related to the weight symmetry commonly observed in most artificial neural networks).


\begin{acknowledgments}
We would like to thank three referees for their insightful comments. We also thank Xiao-Hui Deng and K. Y. Michael Wong for useful discussions. This research was supported by the start-up budget 74130-18831109 of the 100-talent-
program of Sun Yat-sen University (H.H.), and the NSFC (Grant No. 11805284) (H.H.), and grants from the Research Grants Council 
of Hong Kong (16322616 and 16396817) (T.H.).
\end{acknowledgments}
\setcounter{figure}{0}    
\renewcommand{\thefigure}{S\arabic{figure}}
\renewcommand\theequation{S\arabic{equation}}
\setcounter{equation}{0}  

\newpage
\onecolumngrid
\appendix
\section*{Supplemental Material}
\label{methods}
\section{Message Passing Algorithms for unsupervised learning with prior information}
\label{SM-a}
 In our current setting, 
 we assume that the statistical inference of synaptic weights
 from the raw data has the correct prior information $P_{0}(\bx^1,\bx^2)=\prod _{i=1}^{N}P_{0}(\xi_{i}^{1},\xi_{i}^{2})$,
 where $P_{0}(\xi_{i}^{1},\xi_{i}^{2})=\frac{1+q}{4}\delta (\xi^{1}_{i}-\xi^{2}_{i})+\frac{1-q}{4}\delta (\xi^{1}_{i}+\xi^{2}_{i})$, where $q$ 
 is the correlation level between the two receptive fields.
 According to the Bayes' rule, the posterior probability of synaptic weights is given by
\begin{equation}\label{post}
    \begin{split}
        & P(\bm{\xi}^{1},\bm{\xi}^{2}|\{\bs^a\}_{a=1}^{M})=\frac{\prod _{a} P(\bm{\sigma}^{a} |\bm{\xi}^{1},\bm{\xi}^{2} ) \prod_{i=1}^{N} P_{0}({\xi_{i}^{1},\xi_{i}^{2}})}{ \sum_{\bm{\xi}^{1},\bm{\xi}^{2}     }  \prod _{a} P(\bm{\sigma}^{a} |\bm{\xi}^{1},\bm{\xi}^{2}) \prod_{i=1}^{N} P_{0}({\xi_{i}^{1},\xi_{i}^{2}})  }\\
        &= \frac{1}{\Omega} \prod_{a} \frac{1}{\cosh{(\beta^{2}Q)}}\ \cosh{\bigg  (\frac{\beta}{\sqrt{N}}\bm{\xi}^{1}\cdot\bm{\sigma}^{a}  \bigg )}\cosh{ \bigg  (\frac{\beta}{\sqrt{N}}\bm{\xi}^{2}\cdot\bm{\sigma}^{a}\bigg )} \prod_{i=1}^{N}P_{0}(\xi_{i}^{1},\xi_{i}^{2}   ),
    \end{split}
\end{equation}
where $Q\equiv\frac{1}{N}\sum_{i=1}^{N} \xab $, representing the overlap of the two RFs, and $\Omega$  is the so-called partition function in statistical physics.

Using the Bethe approximation~\cite{cavity-2001,Huang-2017,Huang-2019}, we can easily write down the 
belief propagation equations as follows,
\begin{subequations}
\begin{align}
P_{i\rightarrow a}(\xi_{i}^1,\xi_i^2)&=\frac{1}{Z_{i\rightarrow a}}P_{0}(\xi_{i}^{1},\xi_{i}^{2})
\prod_{b\in\partial i\backslash
a}\mu_{b\rightarrow i}(\xi_{i}^1,\xi_i^2), \label{bp0}  \\
\begin{split}
\mu_{b\rightarrow i}(\xi_{i}^1,\xi_i^2)&=\sum_{ \{\bm{\xi}^{1},\bm{\xi}^{2}\} \backslash\{\xi_i^1,\xi_i^2\}}\frac{1}{\cosh\left(\beta^2Q_c+\frac{\beta^2}{N}\xi_i^1\xi_i^2\right)}\cosh\left(\beta X_b+\frac{\beta}{\sqrt{N}}\xi_i^1\sigma_i^b\right)\cosh\left(\beta Y_b+\frac{\beta}{\sqrt{N}}\xi_i^2\sigma_i^b\right)\\
&\times\prod_{j\in\partial
  b\backslash i}P_{j\rightarrow b}(\xi_{j}^1,\xi_j^2),\label{bp1}
\end{split}
\end{align}
\end{subequations}
where we define auxiliary variables $X_{b}=\frac{1}{\sqrt{N}}\sum_{j \neq i}\xi_{j}^{1}\sigma_{j}^{b} $,
$Y_{b}=\frac{1}{\sqrt{N}}\sum_{j \neq i}\xi_{j}^{2}\sigma_{j}^{b} $, and $Q_c=\frac{1}{N}\sum_{j\neq i}\xi_j^1\xi_j^2$. 
$j\in\partial b\backslash i$ indicates neighbors of the data node $b$ excluding the synaptic weight $i$. In our model,
all synaptic weights are used to explain each data sample. The belief propagation is commonly defined in a factor graph representation, where the synaptic-weight-pair
acts as the variable node, while the data sample acts as the factor node (or constraint to be satisfied)~\cite{MM-2009}. The learning can then be interpreted 
as the process of synaptic weight inference based on the data constraints.
The cavity probability $P_{i \to a}(\xi_{i}^{1},\xi_{i}^{2})$
is defined as the probability of the pair $(\xi_{i}^{1},\xi_{i}^{2})$ without considering
the contribution of the data node $a$. $Z_{i\to a}$ is thus a normalization constant for
the cavity probability $P_{i\to a}(\xi_i^1,\xi_i^2)$.
The cavity probability can then be parameterized by the cavity magnetization $m_{i\rightarrow a}^{1,2}$ and correlation $q_{i\rightarrow a}$ as
$ P_{i \to a}(\xi_{i}^{1},\xi_{i}^{2})=\frac{1+m_{i \to a}^{1}\xi_{i}^{1}+m_{i \to a}^{2}\xi_{i}^{2}+q_{i \to a}\xi_{i}^{1}\xi_{i}^{2}}{4}      $.
$\mu_{b\to i}(\xi^1_i,\xi^2_i)$ represents the contribution of one data node given the value of $(\xi^1_i,\xi^2_i)$.
Due to the central limit theorem, $X_b$ and $Y_b$ 
can be considered as two correlated Gaussian random variables.
We thus define $G_{b \to i}^{1}=\frac{1}{\sqrt{N}}\sum_{j \neq i} \sigma_{j}^{b}m_{j \to b}^{1}   $, 
$G_{b \to i}^{2}=\frac{1}{\sqrt{N}}\sum_{j \neq i} \sigma_{j}^{b}m_{j \to b}^{1}$, 
$ \gmba=\frac{1}{N}\sum_{j \neq i}(1-(m^{1}_{j\to b})^{2}) $, and $\gmbb=\frac{1}{N}\sum_{j \neq i}(1-(m^{2}_{j\to b})^{2})$ 
as the means and variances of the two variables, respectively.
The covariance is given by $\Xi_{b \to i}=\frac{1}{N}\sum_{j \neq i}(q_{j \to b}-m_{j \to b}^{1}m_{j \to b}^{2}    )$.
Moreover, $Q_{c}$ is approximated by its cavity mean $Q_{b \to i}= \frac{1}{N} \sum_{j \neq i}q_{j \to b}$.
As a result, the intractable summation in Eq.~(\ref{bp1}) can be replaced by a
jointly-correlated Gaussian integral,
  \begin{equation}\label{mu}
\begin{split}
 \mu_{b\rightarrow i}(\xi_i^1,\xi_i^2)&=\frac{1}{\cosh\left(\beta^2Q_{b\rightarrow i}
 +\frac{\beta^2}{N}\xi^1_i\xi^2_i\right)}\iint DxDy\cosh\left(\beta\sqrt{\gmba}x+\beta\gba+\frac{\beta}{\sqrt{N}}\xi^1_i\sigma_i^b\right)\\
 &\times\cosh\left(\beta\sqrt{\gmbb}(\psi x+\sqrt{1-\psi^2}y)+\beta\gbb+\frac{\beta}{\sqrt{N}}\xi^2_i\sigma_i^b\right),
\end{split}
 \end{equation}
where  the standard Gaussian measure $Dx=\frac{e^{-x^2/2}dx}{\sqrt{2\pi}}$, and $\psi=\frac{\Xi_{b \to i}}{\sqrt{ \Gamma_{b \to i}^{1}  \Gamma_{b \to i}^{2}      }}       $.

We further define the cavity bias $u_{b \to i}(\xi^{1}_{i},\xi^{2}_{i})=\ln{\mu_{b \to i}(\xi^{1}_{i},\xi^{2}_{i}   ) }   $ as
\begin{equation}\label{ub}
\begin{split}
 u_{b\rightarrow i}(\xi_i^1,\xi_i^2)&=\frac{\beta^2}{2}(\gmba+\gmbb+2 \Xi_{b \to i})-\ln\left(2\cosh\Bigl(\beta^2Q_{b\rightarrow i}+\frac{\beta^2\xab}{N}\Bigr)\right)\\
 &+\ln\cosh\left(\beta\gba+\beta\gbb+\frac{\beta}{\sqrt{N}}\sigma_i^b(\xi^1_i+\xi^2_i)\right)\\
 &+
 \ln\left[1+e^{-2\beta^2\Xi_{b \to i}}\frac{\cosh\left(\beta\gba-\beta\gbb+\frac{\beta}{\sqrt{N}}\sigma_i^b(\xi^1_i-\xi^2_i)\right)}{\cosh\left(\beta\gba+\beta\gbb+\frac{\beta}{\sqrt{N}}\sigma_i^b(\xi^1_i+\xi^2_i)\right)}\right].
\end{split}
 \end{equation}
Using Eq.~(\ref{bp0}), the cavity magnetizations $m_{i \to a}^{1}$,  $m_{i \to a}^{2}$,
 and the cavity correlation $ q_{i \to a}  $ can be computed as follows,
\begin{equation}\label{BPm}
    \begin{split}
        & m_{i \to a}^{1}=  \frac{\sum_{\xi_{i}^{1},\xi_{i}^{2}}  \xi_{i}^{1}  e^{ \sum_{b  \in \partial i \backslash a}  u_{b \to i}(\xi_{i}^{1},\xi_{i}^{2})}  P_{0}(\xi_{i}^{1},\xi_{i}^{2})     }{\sum_{\xi_{i}^{1},\xi_{i}^{2}}   e^{ \sum_{b \in  \partial i \backslash a}  u_{b \to i}(\xi_{i}^{1},\xi_{i}^{2})} P_{0}(\xi_{i}^{1},\xi_{i}^{2})    },  \\
        & m_{i \to a}^{2}=  \frac{\sum_{\xi_{i}^{1},\xi_{i}^{2}}  \xi_{i}^{2}  e^{ \sum_{b  \in  \partial i \backslash a}  u_{b \to i}(\xi_{i}^{1},\xi_{i}^{2})}    P_{0}(\xi_{i}^{1},\xi_{i}^{2})     }{\sum_{\xi_{i}^{1},\xi_{i}^{2}}   e^{ \sum_{b\in \partial i \backslash a}  u_{b \to i}(\xi_{i}^{1},\xi_{i}^{2})}  P_{0}(\xi_{i}^{1},\xi_{i}^{2})    },  \\
        & q_{i \to a}=  \frac{\sum_{\xi_{i}^{1},\xi_{i}^{2}} \xi_{i}^{1} \xi_{i}^{2}  e^{ \sum_{b \in \partial i \backslash a}  u_{b \to i}(\xi_{i}^{1},\xi_{i}^{2})}   P_{0}(\xi_{i}^{1},\xi_{i}^{2})     }{\sum_{\xi_{i}^{1},\xi_{i}^{2}}   e^{ \sum_{b  \in\partial i \backslash a}  u_{b \to i}(\xi_{i}^{1},\xi_{i}^{2})}  P_{0}(\xi_{i}^{1},\xi_{i}^{2})    }.
    \end{split}
\end{equation}

Starting from random initialization values of cavity magnetizations and correlations,
the above belief propagation iterates until convergence. 
To carry out the inference of synaptic weights (so-called learning),
one only need to compute the full magnetizations by replacing 
$b  \in \partial i \backslash a$ in Eq.~(\ref{BPm}) 
by $ b  \in \partial i  $. The free energy can also be estimated under the Bethe approximation,
given by $-\beta f_{{\rm Bethe}}=\frac{1}{N}\sum_{i}\Delta f_i-\frac{N-1}{N}\sum_{a}\Delta f_a$ where the single synaptic-weight-pair
contribution $\Delta f_i$ and the single data sample contribution $\Delta f_a$ are given as follows,
\begin{subequations}\label{fifa}
\begin{align}
\Delta f_i&=\ln\sum_{\xi^1_i,\xi^2_i} P_{0}(\xi^{1}_{i},\xi^{2}_{i})   \prod_{b\in\partial i}\mu_{b\rightarrow i}(\xi^1_i,\xi^2_i)   ,\\
\begin{split}
\Delta f_a&=\frac{\beta^2\Gamma_a^2(1-\tilde{\psi}^2)}{2}-\ln\left(2\cosh(\beta^2Q_{a})\right)+\frac{\beta^2}{2}\left(\sqrt{\Gamma_a^1}+\sqrt{\Gamma_a^2}\tilde{\psi}\right)^2\\
 &+\ln\cosh\left(\beta G_{a}^1+\beta G_a^2\right)
 +
 \ln\left[1+e^{-2\beta^2\Xi_a}\frac{\cosh\left(\beta G_a^1-\beta G_a^2\right)}{\cosh\left(\beta G_a^1+\beta G_a^2\right)}\right],
 \end{split}
\end{align}
\end{subequations}
where $G_{a }^{1}=\frac{1}{\sqrt{N}}\sum_{j \in\partial a  } \sigma_{j}^{a}m_{j \to a}^{1}   $, $G_{a }^{2}=\frac{1}{\sqrt{N}}\sum_{j \in\partial a } \sigma_{j}^{a}m_{j \to a}^{2}$, $ \Gamma_{a}^{1}   =\frac{1}{N}\sum_{j \in\partial a}(1-(m^{1}_{j  \to a })^{2}) $, $\Gamma_{a}^{2}=\frac{1}{N}\sum_{j \in\partial a }(1-(m^{2}_{j \to  a })^{2})$, 
$Q_a=\frac{1}{N}\sum_{i\in\partial a}q_{i\to a}$,
$ \Xi_{a} =\frac{1}{N}\sum_{j \in\partial a}\left(q_{j\to a}-m_{j \to a}^1m_{j\to a}^2\right)$, and 
$\tilde{\psi}=\frac{\Xi_{a}}{\sqrt{\Gamma_{a}^{1}\Gamma_{a}^{2}    }}$.
\section{ Replica analysis of the model }
\label{SM-b}
For a replica analysis, we need to evaluate a
disorder average of an integer power of the partition function $ \langle  \Omega^{n}     \rangle    $, 
where $\langle    \bullet     \rangle  $  is the disorder average over the 
true RF distribution $  P_{0}(\bxtF,\bxtS)$ that is factorized over components and the corresponding
data distribution $ P(\{ \bs^{a} \}_{a=1}^{M} |\bxtF,\bxtS)$ as
   \begin{equation}
    \begin{split}
         \langle  \Omega^{n} \rangle  = &\sum_{\{\bxtF,\bxtS,\{\bs^a \}\}  }  \prod_{i=1}^{N}\left[P_{0}(\xi_{i}^{1,{\rm true}},\xi_{i}^{2,{\rm true}} )\right] \prod_{a=1}^{M}  \frac{  \cosh{ \bigg( \frac{\beta}{\sqrt{N}}\bxtF\cdot\bs^{a} \bigg ) } \cosh{ \bigg (  \frac{\beta}{\sqrt{N}}   \bxtS\cdot\bs^{a}  \bigg )  }}{2^{N}e^{\beta^{2}} \cosh{(\beta^{2}q})} \\
        & \times  \sum_{ \{\bm{\xi}^{1,\gamma}, \bm{\xi}^{2,\gamma}   \} } \prod_{a,\gamma}  \frac{\cosh{\bigg( \frac{\beta}{\sqrt{N}} \bm{\xi}^{1,\gamma}\cdot\bm{\sigma}^{a} \bigg )  }   \cosh{\bigg( \frac{\beta}{\sqrt{N}} \bm{\xi}^{2,\gamma}\cdot \bm{\sigma}^{a}   } \bigg ) }{ \cosh{(\beta^{2}R^{\gamma}})  }\prod_{i,\gamma}P_{0}(\xi_{i}^{1,\gamma},\xi_{i}^{2,\gamma}),
    \end{split}  \label{partF}
\end{equation}
where $q=\frac{1}{N}\bxtF\cdot\bxtS$, and $\gamma $ is the replica index.
The typical free energy can then be obtained as $ -\beta f=\lim_{n \to 0, N \to \infty} \frac{\ln{ \langle  \Omega^{n}    \rangle    }}{nN}       $.
To compute explicitly  $  \langle  \Omega^{n}   \rangle $, we need to specify the order parameters as follows:
  \begin{subequations}
  \label{opdef}
    \begin{align}
        &T_{1}^{\gamma}=\frac{1}{N}\bxtF\cdot\bm{\xi}^{1,\gamma},     \quad    T_{2}^{\gamma}=\frac{1}{N}\bxtS \cdot\bm{\xi}^{2,\gamma},
        \quad\tau_{1}^{\gamma}=\frac{1}{N}\bxtF\cdot\bm{\xi}^{2,\gamma}, \quad  \tau_{2}^{\gamma}=\frac{1}{N}\bxtS\cdot \bm{\xi}^{1,\gamma}, \\
        &  q_{1}^{\gamma ,\gamma'}=\frac{1}{N}\bm{\xi}^{1,\gamma}\cdot\bm{\xi}^{1,\gamma'},  \quad q_{2}^{\gamma ,\gamma'}=\frac{1}{N}\bm{\xi}^{2,\gamma}\cdot\bm{\xi}^{2,\gamma'}, 
        \quad  R^{\gamma}=\frac{1}{N}\bm{\xi}^{1,\gamma}\cdot\bm{\xi}^{2,\gamma}, \quad   r^{\gamma  ,\gamma'}=\frac{1}{N} \bm{\xi}^{1,\gamma}\cdot\bm{\xi}^{2,\gamma'}.
    \end{align}
  \end{subequations}
Inserting these definitions in the form of the delta functions as well as
their corresponding integral representations, one can decompose the computation of $\left<\Omega^n\right>$
into entropic and energetic parts. However, to further simplify the computation, we make a simple Ansatz, i.e.,
the order parameters are invariant under the permutation of replica indexes. This is the so-called RS Ansatz.
The RS Ansatz reads,
\begin{subequations}
\label{RS1}
  \begin{align}
      &  R^{\gamma}=R,  \quad i\hat{R} ^{\gamma}=\hat{R},  \quad
        T_{1}^{\gamma}=T_{1},  \quad i\hat{T}_1^{\gamma}=\hat{T_{1}}, \quad
        T_{2}^{\gamma}=T_{2},  \\
     & i\hat{T}_2^{\gamma}=\hat{T_{2}},\quad
      \quad  \tau_{1}^{\gamma}=\tau_{1},  \quad i\hat{\tau}_{1}^{\gamma}=\hat{\tau_{1}}, \quad
        \tau_{2}^{\gamma}=\tau_{2},  \quad  i\hat{\tau}_{2}^{\gamma}=\hat{\tau_{2}},
  \end{align}
\end{subequations}
for any $\gamma$, and
\begin{subequations}
\label{RS2}
  \begin{align}
      &   q_{1}^{\gamma,\gamma'}=q_{1},  \quad    i\hat{q_{1}}^{\gamma,\gamma'}=\hat{q_{1}},  \quad
         q_{2}^{\gamma,\gamma'}=q_{2}, \\
         &    i\hat{q_{2}}^{\gamma,\gamma'}=\hat{q_{2}},   \quad
         r^{\gamma,\gamma'}=r,    \quad   i\hat{r}^{\gamma,\gamma'}=\hat{r}, 
  \end{align}
\end{subequations}
for any  $\gamma$ and $\gamma'$. Note that $(\hat{T}_1,\hat{T}_2,\hat{q}_1,\hat{q}_2,\hat{\tau}_1,\hat{\tau}_2,
\hat{R},\hat{r})$ are conjugated order parameters introduced when using the integral representation of the delta function.

Then we can reorganize $ \langle  \Omega^{n}     \rangle   $ as 
\begin{equation}
\label{sdm}
    \langle   \Omega^{n}  \rangle=\int d \mathcal{O} d\mathcal{\Hat{O}}  e^{N  \mathcal{A} (\mathcal{O},\hat{\mathcal{O}},\alpha ,\beta,q,n) },
\end{equation}
where $\mathcal{O}$ and $\mathcal{\hat{O}}$ denote, respectively, all non-conjugated and conjugated
order parameters.
In the large $N$ limit, the integral is dominated by an equilibrium action:
\begin{equation}
 \label{sdm2}
 \begin{split}
     \mathcal{A}&=- nR\hat{R}-nT_{1}\hat{T_{1}}-nT_{2}\hat{T_{2}}-n\tau_{1}\hat{\tau_{1}}-n\tau_{2}\hat{\tau_{2}}-\frac{n(n-1)}{2}q_{1}\hat{q_{1}}\\
     &-\frac{n(n-1)}{2}q_{2}\hat{q_{2}}-\frac{n(n-1)}{2}r\hat{r}+G_{S}+\alpha G_{E},
     \end{split}
 \end{equation}
where $G_{S}$ is the entropic term, and $G_{E}$ is the energetic term. 

We first compute the entropic term as follows,
\begin{equation}\label{Gs0}
    \begin{split}
        G_{S}&= \ln\left[ \sum_{ \{ \xi^{1,\gamma},\xi^{2,\gamma}  \}}
        \exp\left( \hat{R}\sum_{\gamma=1}^{n}\xi^{1,\gamma}\xi^{2,\gamma}+
        \hat{T_{1}}\sum_{\gamma=1}^n \xi^{1,\gamma}\bxF+
        \hat{T_{2}}\sum_{\gamma=1}^{n}\xi^{2,\gamma}\bxS+
        \hat{\tau_{1}}\sum_{\gamma=1}^n\bxF\xi^{2,\gamma}\right)\right.\\
        &\left.\times\exp\left(\hat{\tau_{2}}\sum_{\gamma=1}^{n} \xi^{1,\gamma}\bxS+ 
        \sum_{\gamma<\gamma'}\left( \hat{q_{1}}\xi^{1,\gamma}\xi^{1,\gamma'}
             +\hat{q_{2}} \xi^{2,\gamma}\xi^{2,\gamma'}
             +\hat{r} \xi^{1,\gamma}\xi^{2,\gamma'}\right)+\sum_{\gamma=1}^{n}\ln P_{0}(\xi^{1,\gamma},\xi^{2,\gamma}) \right)\right]_{\bxF,\bxS}.
    \end{split}         
\end{equation}
After a bit lengthy algebraic manipulation with the techniques developed in our previous work~\cite{Huang-2019},
we arrive at the final result of $G_S$ as
\begin{equation}
\label{gscomp}
    G_{S}= \ln\left[  \int D\mathbf{z}\left ( \sum_{\xi^{1},\xi^{2}}  e^{  b_{1}\xi^{1}+b_{2}\xi^{2}+b_{3}\xi^{1}\xi^{2} +\ln P_{0}(\xi^{1},\xi^{2})  }\right)^{n} \right]_{\bxF,\bxS}-\frac{n}{2}\hat{q}_{1}-\frac{n}{2}\hat{q}_{2},
\end{equation}
where we define $D\mathbf{z}=Dz_1Dz_2Dz_3$ with three independent
standard Gaussian random variables ($z_1$, $z_2$ and $z_3$), $[\bullet]$ denotes a disorder average with respect to
the true prior. From this expression, an effective two-spin interaction Hamiltonian can be extracted, determining the effective partition function $Z_{{\rm eff}}$ in
the main text. The effective fields and coupling are given as follows,
\begin{subequations}
\label{b123}
  \begin{align}
      &  b_{1}=\sqrt{\hat{q_{1}}-\frac{\hat{r}}{2} }z_{1}+\sqrt{\frac{\hat{r}}{2} }z_{3}    +\hat{T_{1}}\bxF+\hat{\tau_{2}}\bxS,\\
      &   b_{2}=\sqrt{ \hat{q_{2}}-\frac{\hat{r}}{2} }z_{2}+\sqrt{\frac{\hat{r}}{2} }z_{3}+\hat{T_{2}}\bxS+\hat{\tau_{1}}\bxF,\\
      & b_{3}=\hat{R}-\frac{\hat{r}}{2}.
  \end{align}
\end{subequations}
Therefore, $Z_{{\rm eff}}=\frac{1+q}{2}e^{b_3}\cosh(b_1+b_2)+\frac{1-q}{2}e^{-b_3}\cosh(b_1-b_2)$.

Next, we compute the energetic term $G_E$ given by
\begin{equation}
\label{gedef}
    G_{E}=\ln \left< \frac{ \cosh{(\beta X^{0})}\cosh{(\beta Y^{0}})     }{e^{\beta^2}\cosh{(\beta^{2}q})}   \prod_{\gamma=1}^{n}  \frac{ \cosh{(\beta X^{\gamma})}\cosh{(\beta Y^{\gamma}} )}{\cosh{(\beta^{2}R^{\gamma})}} \right>, 
\end{equation}
 where $\langle   \bullet \rangle  $ defines the disorder average. $X^{ 0 } =\frac{1}{\sqrt{N}}\sum_{i=1}^{N}\xi_{i}^{1,{\rm true}}\sigma_{i}$, 
 $Y^{ 0 } =\frac{1}{\sqrt{N}}\sum_{i=1}^{N}\xi_{i}^{2,{\rm true}}\sigma_{i}$, and $X^{ \gamma } =\frac{1}{\sqrt{N}}\sum_{i=1}^{N}\xi_{i}^{1,\gamma}\sigma_{i}$, $ Y^{ \gamma } =\frac{1}{\sqrt{N}}\sum_{i=1}^{N}\xi_{i}^{2,\gamma}\sigma_{i} $,
 where $\bs$ represents a typical data sample.
  These four quantities are correlated random Gaussian variables, due to the central limit theorem.
  To satisfy their covariance structure determined by the order parameters,
  the random variables $X^{0},Y^{0},X^{\gamma}$ and $Y^{\gamma} $ are parameterized by six standard Gaussian variables of zero mean and unit variance ($ t_{0},x_{0},u,u',y_{\gamma},\omega_{\gamma} $) as follows,
\begin{subequations}
\label{para}
  \begin{align}
         X^{0}&=t_{0},\\
        Y^{0}&=qt_{0}+\sqrt{1-q^{2}}x_{0},\\
        X^{\gamma}&=T_{1}t_{0}+\frac{\tau_{2}-T_{1}q}{\sqrt{1-q^{2}}}x_{0}+Bu+\sqrt{1-q_{1}}\omega_{\gamma}, \\
      \begin{split}
           Y^{\gamma} &=\tau_{1}t_{0}+\frac{T_{2}-\tau_{1}q}{\sqrt{1-q^{2}}}x_{0}+\frac{r-A}{B}u+\frac{R-r}{\sqrt{1-q_{1}}}\omega_{\gamma}+Ku' \\
           &  +\sqrt{1-q_{2}- \frac{(R-r)^{2}}{1-q_{1}}}y_{\gamma},
      \end{split} 
  \end{align}
\end{subequations}
where $A=T_{1}\tau_{1}+\frac{(\tau_{2}-T_{1}q)(T_{2}-\tau_{1}q)}{1-q^{2}}  $, 
$ B=\sqrt{ q_{1}-(T_{1})^{2}-\frac{(\tau_{2}-T_{1}q)^{2} }{ 1-q^{2}   }          }   $, and $  K=\sqrt{ q_{2}-(\tau_{1})^{2}-\frac{(T_{2}-\tau_{1}q)^{2}}{1-q^{2}}-(\frac{r-A}{B})^{2}      }$.
  Therefore, the $G_{E}$ term can be calculated by a standard Gaussian integration given by
\begin{equation}
\label{GEres}
    \begin{split}
        & G_{E}=\ln   \left [ \int Dt_{0}Dx_{0} Du Du'  \frac{ \cosh{(\beta t^{0})}   \cosh{\beta(qt_{0}+\sqrt{1-q^{2}}x_{0})} }{e^{\beta^2}\cosh{(\beta^{2}q})}\right.\\
        &\left.\times \left( \int D \omega  Dy
         \frac{1}{\cosh{(\beta^{2}R)}}
          \cosh{\beta(T_{1}t_{0}+\frac{\tau_{2}-T_{1}q}{\sqrt{1-q^{2}}}x_{0}+Bu+\sqrt{1-q_{1}}\omega           )}\right.\right.\\
         & \left.\left.\times\cosh\beta(\tau_{1}t_{0}+\frac{T_{2}-\tau_{1}q}{\sqrt{1-q^{2}}}x_{0}+ \frac{r-A}{B} u+ +\frac{R-r}{\sqrt{1-q_{1}}}\omega+Ku'+Cy ) \right)^{n} \right ],
        \end{split}
\end{equation}
where $C\equiv\sqrt{1-q_2-\frac{(R-r)^2}{1-q_1}}$.

By introducing the auxiliary variables as follows,
\begin{subequations}\label{AUV}
  \begin{align}
  Z_{{\rm E}}&=  e^{\beta^{2}(R-r)}\cosh{(\beta  \Lambda_{+})} +  e^{-\beta^{2}(R-r)}\cosh{(\beta  \Lambda_{-})},\\
        \Lambda_{+}&=(T_{1}+\tau_{1})t_{0}+\frac{ (T_{2}+\tau_{2})- q(T_{1}+\tau_{1}) }{ \sqrt{1-q^{2}} }x_{0}+\Bigl(B+\frac{r-A}{B}\Bigr)u+Ku', \\
      \Lambda_{-}&=(T_{1}-\tau_{1})t_{0}+\frac{ (\tau_{2}-T_{2})- q(T_{1}-\tau_{1}) }{ \sqrt{1-q^{2}} }x_{0}+\Bigl(B-\frac{r-A}{B}\Bigr)u-Ku',
  \end{align}
\end{subequations}
we  finally arrive at the free energy $F_{\beta}= -\beta f_{{\rm RS}}$ as
\begin{equation}
\label{freeErep}
    \begin{split}  
    & F_{\beta}= -R\hat{R}-T_{1}\hat{T_{1}}-T_{2}\hat{T_{2}}-\tau_{1}\hat{\tau_{1}}-\tau_{2}\hat{\tau_{2}}+\frac{\hat{q_{1}}}{2}(q_{1}-1)+\frac{\hat{q_{2}}}{2}(q_{2}-1)\\
    &+\frac{r\hat{r}}{2}+\int D\mathbf{z}\left[ \ln Z_{{\rm eff}} \right]_{\bxF,\bxS}
    -\alpha\ln\left(2\cosh(\beta^{2}R)\right)+\alpha \beta^{2}\left(1-\frac{q_{1}+q_{2}}{2}\right)\\
    &+ \frac{  \alpha e^{-\beta^{2}}}{\cosh{(\beta^{2}q})} \int D\mathbf{t}\cosh{\beta t_{0}} \cosh{\beta (qt_{0}+\sqrt{1-q^{2}}x_{0})} \ln Z_{{\rm E}},
    \end{split}
\end{equation}
where $D\mathbf{t}=Dt_0Dx_0DuDu'$. The saddle-point analysis in Eq.~(\ref{sdm}) requires that the order parameters should be the stationary point of
the free energy. All these conjugated and non-conjugated order parameters are subject to saddle-point equations derived from setting the 
corresponding derivatives of the free energy with respect to the order parameters zero. Here we skip the technical details to derive the saddle-point equations.
We refer the interested readers to our previous work~\cite{Huang-2019}.

The saddle-point equations for non-conjugated order parameters are given by
    \begin{subequations}
  \begin{align}\label{SDE-nc}
      T_{1}&=[\bxF\langle \xi^{1} \rangle ], \\
      T_{2}&=[\bxS\langle  \xi^{2} \rangle ],\\
      q_{1}&=[  \langle  \xi^{1}  \rangle^{2}  ], \\
      q_{2}&=[  \langle  \xi^{2}  \rangle^{2}  ],  \\
      \tau_{1}&=[ \bxF \langle  \xi^{2} \rangle ], \\
      \tau_{2}&=[ \bxS \langle  \xi^{1} \rangle ],  \\
      R&=[\langle \xi^{1}\xi^{2} \rangle], \\
      r&=[\langle \xi^{1}\rangle \langle \xi^{2} \rangle ],
  \end{align}
\end{subequations}
where $[\bullet]$ indicates an average over the standard Gaussian
random variables $\mathbf{z}$ and the true prior $P_0(\bxF,\bxS)$,
and $ \langle   \bullet  \rangle  $ is an average under the effective Boltzmann
distribution $P(\xi^{1},\xi^{2})=\frac{1}{Z_{{\rm eff}}}e^{b_{1}\xi^{1}+b_{2}\xi^{2}+b_{3}\xi^{1}\xi^{2}+\ln{P_{0}(\xi^{1},\xi^{2})}} $.

It is straightforward to show that
\begin{equation}
    \begin{split}
      &\langle \xi^{1} \rangle= \frac{\partial}{\partial b_{1}} \ln Z_{{\rm eff}}  \\
     &=\frac{\tanh b_{1}(1+q\tanh b_3)+\tanh b_{2}(q+\tanh b_{3})}{1+q\tanh b_{3}+\tanh{b_{1}}\tanh{b_{2}}(q+\tanh{b_{3}})}.
    \end{split}  \label{eqxia} 
\end{equation}
The expression of $\left<\xi^2\right>$ is obtained by exchange of $b_1$ and $b_2$ in Eq.~(\ref{eqxia}).
The correlation term is computed similarly,
\begin{equation}
    \begin{split}
     \langle \xi^{1} \xi^{2} \rangle &= \frac{\partial}{\partial b_{3}} \ln Z_{{\rm eff}}    \\
     &=\frac{q+\tanh{b_{3}}+\tanh{b_{1}}\tanh{b_{2}}(1+q\tanh{b_{3}})}{1+\tanh{b_{1}}\tanh{b_{2}}(q+\tanh{b_{3}})+  q\tanh{b_{3}}}.
    \end{split}  \label{eqcorre}
\end{equation}

The saddle-point equations for conjugated order parameters are given by
\begin{subequations}\label{rbmReplica3}
\begin{align}
\hat{T}_1&=\alpha\beta^2\langle\langle G_s^+\rangle\rangle,\\
\hat{T}_2&=\alpha\beta^2\langle\langle\langle G_s^-\rangle\rangle\rangle,\\
\hat{q}_1&=\alpha\beta^2\left<(G_s^+)^2\right>,\\
\hat{q}_2&=\alpha\beta^2\left<(G_s^-)^2\right>,\\
\hat{\tau}_1&=\alpha\beta^2\langle\langle G_s^-\rangle\rangle,\\
\hat{\tau}_2&=\alpha\beta^2\langle\langle\langle G_s^+\rangle\rangle\rangle,\\
\hat{R}&=\alpha\beta^2\left<G_c^-\right>-\alpha\beta^2\tanh(\beta^2R),\\
\hat{r}&=2\alpha\beta^2\left<G_s^+G_s^-\right>,
\end{align}
\end{subequations}
where we define three different measures as
\begin{subequations}
 \begin{align}
  \left<\bullet\right> &\equiv\frac{e^{-\beta^2}}{\cosh(\beta^2q)}\int D\mathbf{t}\cosh(\beta t_0)\cosh(\beta qt_0+\beta\sqrt{1-q^2}x_0)\bullet,\\
  \langle\langle\bullet\rangle\rangle &\equiv\frac{e^{-\beta^2}}{\cosh(\beta^2q)}\int D\mathbf{t}\sinh(\beta t_0)\cosh(\beta qt_0+\beta\sqrt{1-q^2}x_0)\bullet,\\
  \langle\langle\langle\bullet\rangle\rangle\rangle &\equiv\frac{e^{-\beta^2}}{\cosh(\beta^2q)}\int D\mathbf{t}\cosh(\beta t_0)\sinh(\beta qt_0+\beta\sqrt{1-q^2}x_0)\bullet,
 \end{align}
\end{subequations}
and three auxiliary quantities as
\begin{subequations}\label{rbmReplica4}
\begin{align}
G_c^-&=\frac{e^{\beta^2(R-r)}\cosh\beta\Lambda_+-e^{-\beta^2(R-r)}\cosh\beta\Lambda_{-}}{e^{\beta^2(R-r)}\cosh\beta\Lambda_++e^{-\beta^2(R-r)}\cosh\beta\Lambda_{-}},\\
G_s^+&=\frac{e^{\beta^2(R-r)}\sinh\beta\Lambda_++e^{-\beta^2(R-r)}\sinh\beta\Lambda_{-}}{e^{\beta^2(R-r)}\cosh\beta\Lambda_++e^{-\beta^2(R-r)}\cosh\beta\Lambda_{-}},\\
G_s^-&=\frac{e^{\beta^2(R-r)}\sinh\beta\Lambda_+-e^{-\beta^2(R-r)}\sinh\beta\Lambda_{-}}{e^{\beta^2(R-r)}\cosh\beta\Lambda_++e^{-\beta^2(R-r)}\cosh\beta\Lambda_{-}}.
\end{align}
\end{subequations}

\section{A small-$q$ expansion of the order parameter $r$}
To understand why the order parameter $r$ with prior is higher than that without prior, we carry out a small-$q$ expansion of the order parameter.
In the small-$q$ limit, we can reasonably assume that the conjugated parameters are close to those estimated under the prior-free case, thereby avoiding a complex analysis of the original iterative equations.
We first denote some short-hand notations, as $t_1\equiv\tanh b_1$, $t_2\equiv\tanh b_2$, $t_3\equiv\tanh b_3$, $a\equiv t_1+t_2t_3$, $b\equiv t_2+t_1t_3$, and
$c\equiv 1+t_1t_2t_3$. It then follows that
\begin{subequations}\label{Sq1}
 \begin{align}
  \left<\xi^1\right>_p\left<\xi^2\right>_p &=\frac{(t_1(1+qt_3)+t_2(q+t_3))(t_2(1+qt_3)+t_1(q+t_3))}{(1+qt_3+t_1t_2(q+t_3))^2}\\
  &=\frac{\left(\left<\xi^1\right>_{pf}+q\left<\xi^2\right>_{pf}\right)\left(\left<\xi^2\right>_{pf}+q\left<\xi^1\right>_{pf}\right)}{\left(1+q\left<\xi^1\xi^2\right>_{pf}\right)^2}\\
  &=\left<\xi^1\right>_{pf}\left<\xi^2\right>_{pf}+q\left(\left<\xi^1\right>_{pf}^2+\left<\xi^2\right>_{pf}^2\right)-2q\left(\left<\xi^1\xi^2\right>_{pf}\left<\xi^1\right>_{pf}\left<\xi^2\right>_{pf}\right)+\mathcal{O}(q^2).
  \end{align}
\end{subequations}
where we have used the result of the prior-free case, i.e., $\left<\xi^1\right>_{pf}=\frac{a}{c}$, $\left<\xi^2\right>_{pf}=\frac{b}{c}$, and
$\left<\xi^1\xi^2\right>_{pf}=\frac{t_3+t_1t_2}{c}$. The subscripts p and pf indicate prior and prior-free, respectively. After taking the disorder average, i.e., $[\bullet]$,
we obtain the following relationship:
\begin{equation}\label{Sq2}
 r_p=r_{pf}+q\left([ \langle  \xi^{1}  \rangle^{2}_{pf}]+[ \langle  \xi^{2}  \rangle^{2}_{pf}]-2[\langle \xi^{1}\rangle_{pf} \langle \xi^{2} \rangle_{pf}\langle \xi^{1} \xi^{2} \rangle_{pf}]\right).
\end{equation}
For positive $q$, the last term is verified to be non-negative and decreasing as $\alpha$ increases. For negative $q$, the result also holds. Therefore the last term contributes to the larger value of
the order parameter $r_p$ (more precisely, in absolute value) compared with the prior-free case.

\section{Derivation of the critical point $\alpha_{c}$ for spontaneous symmetry breaking  }
When $\alpha$ approaches the SSB threshold from below, all order parameters get close to zero, except for
  $R$ which is always equal to $q$ due to the prior information. It is easy to show that $\hat{R}$ is also zero below the SSB threshold.
  Therefore, $b_1$, $b_2$ and $b_3$ are all small quantities. Then we can expand our order parameters to leading order.
  Note that $\langle  \xi^{1} \rangle\simeq  b_{1}+qb_{2} $, and
  $\langle  \xi^{2} \rangle  \simeq b_{2}+qb_{1} $. It then follows that
  \begin{subequations}
    \begin{align}\label{apprOP}
      T_{1}&= [\bxF \langle  \xi^{1}  \rangle   ]\simeq\Hat{T_{1}}+q\Hat{\tau_{2}}+q\hat{\tau_{1}}+q^{2}\hat{T_{2}},\\
      T_{2}&= [\bxS \langle  \xi^{2}  \rangle   ]\simeq\Hat{T_{2}}+q\Hat{\tau_{2}}+q\hat{\tau_{1}}+q^{2}\hat{T_{1}},\\
      \tau_{1}&= [\bxF \langle  \xi^{2}  \rangle]\simeq\Hat{\tau_{1}}+q \hat{T_{1}}+q\hat{T_{2}}+q^{2}\hat{\tau_{2}},\\
      \tau_{2}&= [\bxS \langle  \xi^{1}  \rangle]\simeq\Hat{\tau_{2}}+q \hat{T_{1}}+q\hat{T_{2}}+q^{2}\hat{\tau_{1}}.
       \end{align}
  \end{subequations}

Because $R=q$, by defining $W(q)=\frac{e^{\beta^2q}}{2\cosh(\beta^2q)}$, one arrives at the approximation
$G_s^{\pm}\simeq\beta W(q)(\Lambda_+\mp\Lambda_{-})\pm\beta\Lambda_{-}$. To proceed, it is worth noticing that
\begin{subequations}
  \begin{align}
      & \langle  \langle \Lambda_{+}   \rangle   \rangle=\beta[T_{1}+\tau_{1}+\tau_2\tanh{(\beta^{2}q)}+T_2\tanh{(\beta^{2}q)} ],\\
        &  \langle  \langle \Lambda_{-}   \rangle   \rangle=\beta[T_{1}-\tau_{1}+\tau_2\tanh{(\beta^{2}q)}-T_2\tanh{(\beta^{2}q)} ],\\
        & \langle \langle \langle \Lambda_{+}   \rangle \rangle  \rangle=\beta[T_{2}+\tau_{2}+\tau_1\tanh{(\beta^{2}q)}+T_1\tanh{(\beta^{2}q)} ],\\
        & \langle \langle \langle \Lambda_{-}   \rangle \rangle
        \rangle=\beta[\tau_{2}-T_{2}+T_1\tanh{(\beta^{2}q)}-\tau_1\tanh{(\beta^{2}q)} ].\\
  \end{align}
\end{subequations}
Based on the above approximations, it is easy to derive the following approximate values of the relevant conjugated quantities
\begin{subequations}
  \begin{align}
      &  \hat{T}_{1}     \simeq     \alpha \beta^{4}[T_{1}+\Upsilon \tau_{1}+\tau_2\tanh{(\beta^{2}q)}+\Upsilon T_2\tanh{(\beta^{2}q)} ],\\
        &  \hat{T}_{2}  \simeq      \alpha \beta^{4}[T_{2}+\Upsilon \tau_{2}+\tau_1\tanh{(\beta^{2}q)}+\Upsilon T_1\tanh{(\beta^{2}q)} ],\\
        &  \hat{\tau}_{1}   \simeq            \alpha \beta^{4}[\tau_{1}+ \Upsilon T_{1}+T_2\tanh{(\beta^{2}q)}+\Upsilon\tau_2\tanh{(\beta^{2}q)} ],\\
        &  \hat{\tau}_{2}  \simeq     \alpha \beta^{4}[\tau_{2}+\Upsilon T_{2}+T_1\tanh{(\beta^{2}q)}+\Upsilon\tau_1\tanh{(\beta^{2}q)} ],
  \end{align}
\end{subequations}
where $ \Upsilon\equiv2W(q)-1  $.

\begin{figure}
     \includegraphics[bb=10 201 562 579,scale=0.55]{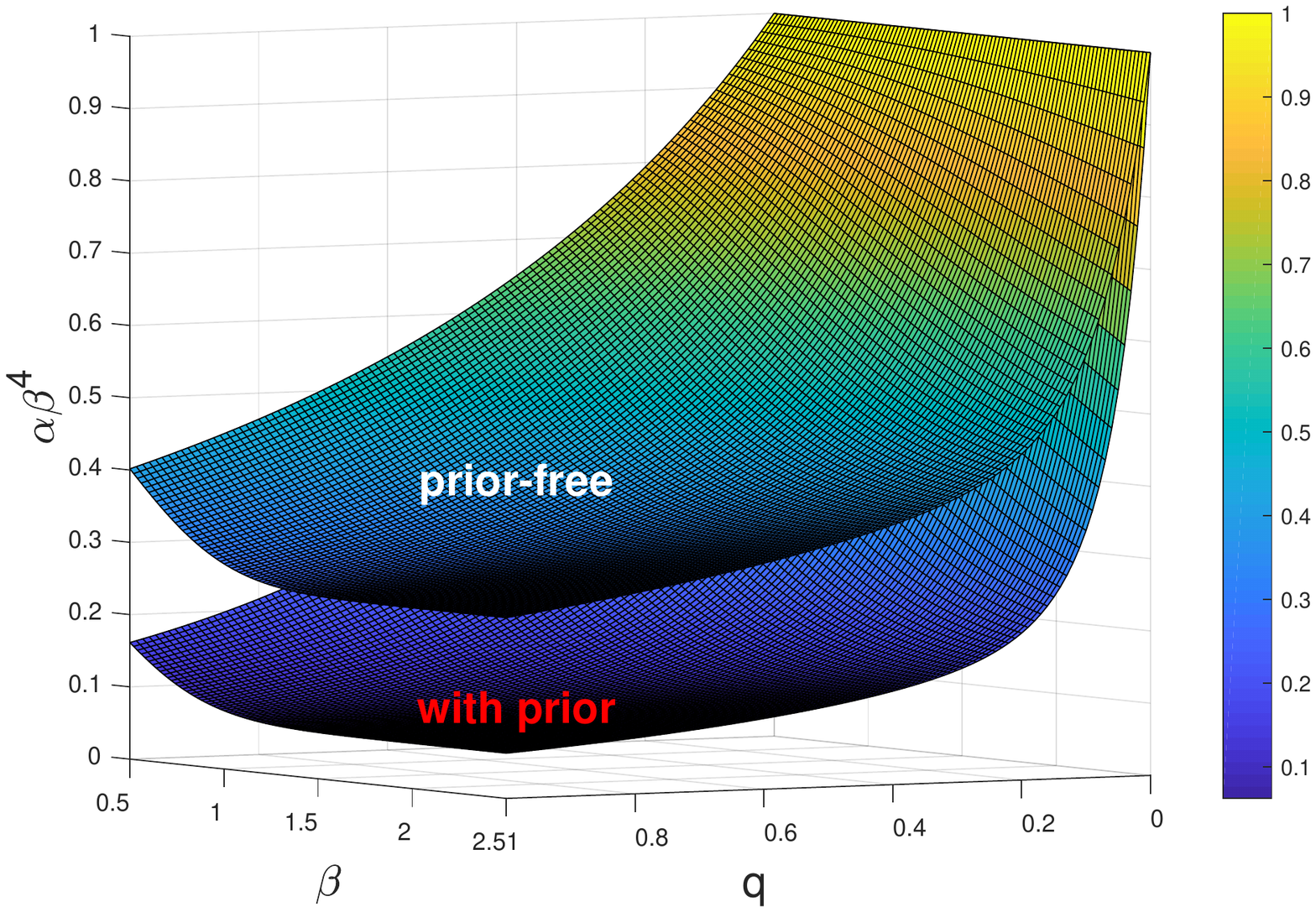}
  \caption{
  (Color online) Comparison of SSB critical data densities in models with/without prior knowledge.
  }\label{comp}
\end{figure}

The above approximations of ($T_1,T_2,\tau_1,\tau_2$) and ($\hat{T}_1,\hat{T}_2,\hat{\tau}_1,\hat{\tau}_2$) can be easily recast into a compact
matrix form as follows,
\begin{equation}
  \left(
\begin{array}
{c}
T_{1}\\
T_{2}\\
\tau_{1}\\
\tau_{2} 

\end{array}
\right)=\left(
\begin{array}
{cccc}
1 & q^{2} &q &q\\
q^{2} & 1  &q &q \\
q & q & 1  & q^{2} \\
q & q & q^{2} & 1
\end{array}
\right)\left(
\begin{array}
{c}
\Hat{T}_{1}\\
\hat{T_{2}}\\
\hat{\tau_{1}}\\
\hat{\tau_{2}}
\end{array}
\right),  \label{matrix1}
\end{equation}
and
\begin{equation}
 \left(
\begin{array}
{c}
\Hat{T}_{1}\\
\hat{T_{2}}\\
\Hat{\tau}_{1}\\
\hat{\tau_{2}}
\end{array}
\right)=\alpha \beta^{4}  
 \left(
\begin{array}
{cccc}
1&  \Upsilon \tanh{(\beta^{2}q)} & \Upsilon &     \tanh{(\beta^{2}q)}\\
\Upsilon \tanh{(\beta^{2}q)}  & 1 & \tanh{(\beta^{2}q)} & \Upsilon \\
\Upsilon & \tanh{(\beta^{2}q)} &1 & \Upsilon \tanh{(\beta^{2}q)} \\
\tanh{(\beta^{2}q)} & \Upsilon & \Upsilon \tanh{(\beta^{2}q)} &1
\end{array}
\right)\left(
\begin{array}
{c}
T_{1}\\
T_{2} \\
\tau_{1} \\
\tau_{2}
\end{array}
\right).\label{matrix2}
\end{equation}
A linear stability analysis implies that the stability matrix $\mathcal{M}$ can be organized in this case as
a block matrix of the form $ \mathcal{M}=  \left( \begin{array}
{cc}
A  &  B\\
B  & A
\end{array}
\right) $, where the matrices $A$ and $B$ are derived from Eq.~(\ref{matrix1}) and Eq.~(\ref{matrix2}), and given respectively by
\begin{subequations}
  \begin{align}
      & A=\alpha \beta^{4} \left(
\begin{array}
{cc}
(1+q\tanh{(\beta^{2}q)})(1+q\Upsilon)  &  (\tanh{(\beta^{2}q)}+q)(q+\Upsilon)   \\
(\tanh{(\beta^{2}q)}+q)(\Upsilon+q) & (1+q\tanh{(\beta^{2}q)})(1+q\Upsilon)
\end{array}
\right),  \\
&B=\alpha \beta^{4} \left(
\begin{array}
{cc}
 (\Upsilon+q)(1+q\tanh{(\beta^{2}q)}) & (\Upsilon q+1)(q+\tanh{(\beta^{2}q)})    \\
 (\Upsilon q+1)(q+\tanh{(\beta^{2}q)})  & (\Upsilon+q)(1+q\tanh{(\beta^{2}q)})
\end{array}
\right).
  \end{align}
\end{subequations}
According to the determinant identity for a block matrix, $ \lvert  \mathcal{M}-\lambda I   \rvert=$  $ \lvert  A+B-\lambda I  \rvert  $    $  \lvert A-B-\lambda I   \rvert $,
the eigenvalues of the stability matrix can be determined by the following two equations,
\begin{equation}
     \left  \lvert
\begin{array}
{cc}
\alpha \beta^{4}(1+q)(1+q\tanh{(\beta^{2}q)} )(1+\Upsilon)-\lambda   & \alpha \beta^{4}(1+q)(\Upsilon+ 1)(q+\tanh(\beta^{2}q))   \\
  \alpha \beta^{4}(1+q)(\Upsilon+ 1)(q+\tanh{(\beta^{2}q)})     &  \alpha \beta^{4}(1+q)(1+q\tanh{(\beta^{2}q)})(1+\Upsilon)-\lambda
\end{array}
\right  \rvert =0,
\end{equation}
and
\begin{equation}
    \left  \lvert   
\begin{array}
{cc}
\alpha \beta^{4}(1-q)(1+q\tanh{(\beta^{2}q)})(1-\Upsilon)-\lambda   & \alpha \beta^{4}(1-q)(\Upsilon-1)(q+\tanh{(\beta^{2}q)})   \\
  \alpha \beta^{4}(1-q)(\Upsilon-1)(q+\tanh{(\beta^{2}q)})     &  \alpha \beta^{4}(1-q)(1+q\tanh{(\beta^{2}q)})(1-\Upsilon)-\lambda
\end{array}
 \right  \rvert =0.
\end{equation}
Using the mathematical identity $\max(1-q,1+q)=1+|q|$, and $\max(1-\Upsilon,1+\Upsilon)=1+|\Upsilon|$,
we conclude that the maximal value of all eigenvalues is given by $ \lambda_{{\rm max}}=\alpha \beta^{4} (1+|q|)(1+|\Upsilon|)(1+q\tanh{(\beta^{2}q)}+|q+\tanh{(\beta^{2}q)}|)$.
The critical data density for the SSB phase is thus given by
\begin{equation}\label{thr-SM}
    \alpha_{c}=\frac{\beta^{-4}}{ (1+|q|)(1+|\Upsilon|)(1+q\tanh{(\beta^{2}q)}+|q+\tanh{(\beta^{2}q)}|)     }.
\end{equation}
This SSB critical data density is compared with that of the prior-free case (i.e., $\Lambda(\beta,q)$ in the main text, see also our previous work~\cite{Huang-2019}) in Fig.~\ref{comp}.
We clearly see that the prior knowledge about $q$ significantly reshapes the critical data density surface for the SSB phase, which provides deep insights about roles of prior information beyond our
previous work~\cite{Huang-2019}.

\section{   Prediction of noise level $\beta$ and correlation level $q$  from raw data}
We first write the posterior probability of the hyper-parameters $\beta$ and $q$ as
\begin{equation}\label{hyperP}
    P(\beta,q|\mathcal{D})=\sum_{\bm{\xi}^{1},\bm{\xi}^{2}}P(\beta,q,  \bm{\xi}^{1},\bm{\xi}^{2}|\mathcal{D} )=\sum_{\bm{\xi}^{1},\bm{\xi}^{2}}   \frac{ P(\mathcal{D}| \beta,q ,\bm{\xi}^{1},\bm{\xi}^{2}    )P_{0}(\bm{\xi}^{1},\bm{\xi}^{2}|q) }{\int  \int d \beta  d q  \sum_{\bm{\xi}^{1},\bm{\xi}^{2}} P(\mathcal{D} |\beta,q , \bm{\xi}^{1},\bm{\xi}^{2}) P_{0}(\bm{\xi}^{1},\bm{\xi}^{2}|q)  },
\end{equation}
where we have used the Bayes' rule, and we assume that $P_0(\bx^1,\bx^2,\beta,q)=P_0(\bx^1,\bx^2|q)\tilde{P}_0(\beta,q)$ where $\tilde{P}_0(\beta,q)$ is a constant or we have no prior knowledge about the true values of
the hyper-parameters. Therefore, we have
\begin{equation}\label{hyperP2}
    P(\beta, q|\mathcal{D} ) \propto \sum_{\bm{\xi}^{1},\bm{\xi}^{2}   }  \prod_{a=1}^{M}P(\bm{\sigma}^{a}|\beta,q,\bx^1,\bx^2)\prod_{i=1}^{N} P_{0}(\xi_{i}^{1},\xi_{i}^{2}|q).
\end{equation}
Note that the data distribution can be expressed as
\begin{equation}\label{dataP}
     P(\bm{\sigma}^{a}| \beta,q, \bx^1,\bx^2 )= \frac{\cosh{\bigg(\frac{\beta}{\sqrt{N}} \bx^{1}\cdot\bs^{a}  \bigg) } \cosh{\bigg(\frac{\beta}{\sqrt{N}} \bm{\xi}^{2}\cdot\bs^{a} \bigg)}}{2^Ne^{\beta^2}\cosh{(\beta^{2}Q)}}. 
\end{equation}
The posterior probability of the hyper-parameters can be finally simplified as $P(\beta,q|\mathcal{D})\propto e^{-\beta^2M}\Omega$, where $\Omega$ is exactly
the partition function of the posterior $P(\bx^1,\bx^2|\mathcal{D})$. This partition function can be written explicitly as follows,
\begin{equation}
   \Omega(\beta,q)=\sum_{\bm{\xi}^{1},\bm{\xi}^{2}} \prod_{a=1}^{M} \frac{\cosh{\bigg(\frac{\beta}{\sqrt{N}}  \bm{\xi}^{1}\cdot\bm{\sigma}^{a}\bigg)}  \cosh{\bigg(\frac{\beta}{\sqrt{N}}  \bm{\xi}^{2}\cdot\bm{\sigma}^{a} \bigg)}}{\cosh{(\beta^{2}Q})} \prod_{i=1}^{N}P_{0}(\xi_{i}^{1},\xi_{i}^{2}|q). 
\end{equation}

Searching for consistent hyper-parameters ($\beta,q$) compatible with the supplied dataset is equivalent to
maximizing the posterior $P(\beta,q|\mathcal{D})$. Following this principle, we first derive the temperature equation as
\begin{equation}\label{Teq}
    \frac{\partial   \ln P(\beta,q |\mathcal{D})     }{\partial \beta}=-2M\beta+\frac{\partial }{\partial  \beta} \ln  \Omega(\beta,q).
\end{equation}
Note that in statistical physics, the energy function is given by $N\epsilon=-\frac{\partial\ln\Omega}{\partial\beta}$, where $\epsilon(\beta,q)$ denotes the energy density (per neuron).
We thus conclude that $\beta$ should obey the following temperature equation,
\begin{equation}\label{betaEq}
    \beta=-\frac{ \epsilon(\beta,q ) }{2\alpha}.
\end{equation}
Note that when the true prior is taken into account, the energy density of the model is analytic with the result $\epsilon=-2\alpha\beta$ independent of $q$.

Given the dataset and an initial guess of $\beta$, the aforementioned message passing scheme can be used to
estimate the energy density of the system as $ N \epsilon=-\sum_{i}\Delta \epsilon_{i}+(N-1)\sum_{a} \Delta \epsilon_{a}$ based on the Bethe approximation. The energy contribution of one synapse-pair reads
\begin{equation}
    \Delta \epsilon_{i}=\frac{\sum_{\xi_{i}^{1},\xi_{i}^{2}} \sum_{b \in\partial i}\frac{\partial u_{b \to i}(\xi_{i}^{1},\xi_{i}^{2})   }{\partial \beta    } e^{\sum_{b\in\partial i}u_{b \to i}(\xi_{i}^{1},\xi_{i}^2)+\ln P_{0}(\xi_{i}^1,\xi^2_i) }   }{\sum_{\xi_{i}^{1},\xi_{i}^{2}   } e^{ \sum_{b \in \partial i} u_{b \to i}(\xi_{i}^{1},\xi_{i}^{2})+\ln P_{0}(\xi_{i}^{1},\xi_{i}^{2})} },
\end{equation}
where $   \frac{\partial u_{b \to i}(\xi_{i}^{1},\xi_{i}^{2})  }{\partial \beta}     $  reads as follows,
\begin{equation}\label{pub}
    \begin{split}
      &\beta\frac{\partial u_{b \to i}(\xi_{i}^{1},\xi_{i}^{2})  }{\partial \beta}  =\beta^2 [\Gamma^{1}_{b \to i}+\Gamma^{1}_{b \to i}+2\Xi_{b \to i}]        -2\beta^2\left(Q_{b \to i}+\frac{\xi_{i}^{1}\xi_{i}^{2}}{N}\right)\tanh{\bigg (\beta^{2}Q_{b \to i}+\frac{\beta^{2}}{N}\xi^{1}_{i}\xi^{2}_{i}   \bigg )}\\
      &+Y_{b\to i}\tanh Y_{b\to i}+\frac{\Delta_{b\to i}}{1+\Delta_{b\to i}}\left(-4\beta^2\Xb+X_{b\to i}\tanh X_{b\to i}-Y_{b\to i}\tanh Y_{b\to i}\right),
     \end{split}
\end{equation}
where $X_{b\to i}\equiv\beta\gba-\beta\gbb+\frac{\beta}{\sqrt{N}}\sigma_i^b(\xi^1_i-\xi^2_i)$, $Y_{b\to i}\equiv\beta\gba+\beta\gbb+\frac{\beta}{\sqrt{N}}\sigma_i^b(\xi^1_i+\xi^2_i)$,
and $\Delta_{b\to i}\equiv e^{-2\beta^2\Xb}\frac{\cosh X_{b\to i}}{\cosh Y_{b\to i}}$. The energy contribution of one data sample is given by
\begin{equation}
    \begin{split}
        &\beta\Delta  \epsilon_{a}=\beta^2 (\Gamma^{1}_{a}+\Gamma_{a}^{2} +2\Xi_{a} )-2\beta^2 Q_{a} \tanh{(\beta^{2}Q_{a})}+Y_a\tanh Y_a\\
        &+\frac{\Delta_a}{1+\Delta_a}\left(-4\beta^2\Xi_a+X_a\tanh X_a-Y_a\tanh Y_a\right),
    \end{split}
\end{equation}
where $X_a\equiv\beta G_a^1-\beta G_a^2$, $Y_a\equiv\beta G_a^1+\beta G_a^2$, and $\Delta_a=e^{-2\beta^2\Xi_a}\frac{\cosh X_a}{\cosh Y_a}$.

Next, we derive the correlation equation. Note that $ P_{0}(\xi_{i}^{1},\xi_{i}^{2}) =\frac{ e^{J_{0}\xi_{i}^{1}\xi_{i}^{2}}}{4\cosh{J_{0}}} $, where $ J_{0}=\tanh^{-1}{q}   $.
We then have 
\begin{equation}
 \frac{\partial P(\beta,q|\mathcal{D})}{\partial q}=e^{-M\beta^2}\frac{\partial\Omega}{\partial q}=0,
\end{equation}
which requires that $\frac{\partial\Omega}{\partial q}=0$. It then follows that
\begin{equation}\label{Domeg}
\begin{split}
 \frac{\partial\Omega}{\partial q}&=\Omega\sum_{\bx^1,\bx^2}P(\bx^1,\bx^2|\mathcal{D})\sum_{i}(\xab-\tanh J_0)\frac{\partial J_0}{\partial q}\\
 &=\Omega\left(\sum_i\left<\xab\right>_{P(\bx^1,\bx^2|\mathcal{D})}-N\tanh J_0\right)\frac{\partial J_0}{\partial q}=0.
 \end{split}
\end{equation}
To satisfy Eq.~(\ref{Domeg}), we must enforce the following correlation equation,
\begin{equation}\label{qEq}
    q=\frac{1}{N}\sum_{i} q_{i},
\end{equation}
where $q_i$ can be computed in a single instance by iterating the message passing scheme.
More precisely,
\begin{equation}\label{qEq02}
 q_{i}=  \frac{\sum_{\xi_{i}^{1},\xi_{i}^{2}} \xi_{i}^{1} \xi_{i}^{2}  e^{ \sum_{b \in \partial i}  u_{b \to i}(\xi_{i}^{1},\xi_{i}^{2})}   P_{0}(\xi_{i}^{1},\xi_{i}^{2})     }{\sum_{\xi_{i}^{1},\xi_{i}^{2}}   e^{ \sum_{b  \in\partial i }  u_{b \to i}(\xi_{i}^{1},\xi_{i}^{2})}  P_{0}(\xi_{i}^{1},\xi_{i}^{2})    }.
\end{equation}

In addition, the negative log-likelihood of the hyper-parameter posterior per neuron can be estimated as $\frac{\mathcal{L}}{N}=\mathcal{C}-\frac{\ln\Omega}{N}+\alpha\beta^2$,
where $\mathcal{C}$ is an irrelevant constant, and the second term can be approximated by $\beta f_{{\rm Bethe}}$.

We finally remark that, in the original Nishimori model~\cite{Nishimori-1980}, a parameter $p$ 
is used to characterize the coupling bias in a multi-spin interaction model (e.g., two-body interactions), while in our case $q$
is used to characterize the correlation level among synapses (couplings). Therefore, $p$ and $q$ are physically different. 
In addition, the Nishimori line is specified by $T=T(p)$, where $T$ is the temperature of the model, and $T(p)$
is a temperature-like parameter expressed as a function of $p$~\cite{Nishimori-1980}. 
However, in our setting, the energy is analytic, i.e., $\epsilon=-2\alpha\beta$, only when the student uses the same true $q$ and the same temperature as the true one 
(\textit{not} a $q$-transformed temperature).
During the expectation-maximization estimation, the estimated $q$ (or $\beta$) is a complex function of $\beta$ and $q$ at a previous step (see Eqs.~(\ref{betaEq}),~(\ref{qEq}) and~(\ref{qEq02})), therefore
$\beta$ and $q$ have a highly non-trivial relationship, unlike simply $T=T(p)$ in the original Nishimori model. 



\end{document}